%% file: draft.tex
\RequirePackage{lineno}
\documentclass[aps,prl,twocolumn,superscriptaddress,preprintnumbers,amsmath,amssymb]{revtex4}

\usepackage{graphicx} 
\usepackage{dcolumn}  
\usepackage{rotating}
\usepackage{booktabs}
\usepackage{SIunits} 
\usepackage{graphicx}
\usepackage{dcolumn}
\usepackage{diagbox}
\usepackage{bm}
\usepackage{color}
\usepackage[english]{babel}
\usepackage{titlesec}
\usepackage{multirow}
\usepackage[colorlinks,linkcolor=blue,urlcolor=blue,anchorcolor=green,citecolor=blue,breaklinks=true]{hyperref}
\usepackage{epstopdf}
\usepackage{enumitem}
\usepackage{hhline}
\usepackage{array}
\usepackage{threeparttable}
\usepackage{ulem}
\usepackage{url}
\usepackage{hyperref}
\usepackage[hyphenbreaks]{breakurl}
\usepackage{orcidlink} 

\newcommand{\BR}{{\cal B}}

\newcommand{\piz}{\pi^{0}}
\newcommand{\EE}{e^+e^-}
\newcommand{\Dsz}{D^{*}_{s0}(2317)^{+}}
\newcommand{\Dso}{D_{s1}(2460)^{+}}
\newcommand{\Dss}{D_{s}^{*+}}
\newcommand{\Ds}{D_{s}^{+}}
\newcommand{\Rdd}{ D_{s}^{*+} \gamma}
\newcommand{\Hdd}{ D_{s}^{+} \pi^{0}}
\newcommand{\Rd}{D^{*}_{s0}(2317)^{+} \to D_{s}^{*+} \gamma}
\newcommand{\Hd}{D^{*}_{s0}(2317)^{+} \to D_{s}^{+} \pi^{0}}
\newcommand{\BrR}{\BR(\Rd)/\BR(\Hd)}
\newcommand{\jp}{J^{P}}

\begin{document}
\hyphenpenalty=10000
\tolerance=1000
\vspace*{-3\baselineskip}
		

\title{\boldmath Observation of the radiative decay $\Rd$}
\input{pub100-orcid}

\begin{abstract}
We report the first observation of the radiative decay $\Rd$ with a statistical significance exceeding 10 standard deviations. The signal is observed in the continuum $e^+ e^- \to c\bar{c}$ process, using combined data samples of 980.4~$\rm fb^{-1}$ from Belle and 427.9~$\rm fb^{-1}$ from Belle II, collected at the KEKB and SuperKEKB asymmetric-energy $\EE$ colliders, respectively. The branching fraction ratio $\BrR$ is measured to be $[7.13 \pm 0.70({\rm stat.}) \pm 0.26({\rm syst.})]\%$.  This result provides crucial discrimination between theoretical models of the $\Dsz$ structure.

\end{abstract}

\maketitle
	
\tighten
	
{\renewcommand{\thefootnote}{\fnsymbol{footnote}}}
\setcounter{footnote}{0}
	
The study of exotic hadrons represents a pivotal frontier in particle physics, providing critical insights into the non-perturbative regime of quantum chromodynamics~\cite{Brambilla:2019esw,Chen:2016spr,Guo:2017jvc}. The scalar charm-strange meson $\Dsz$ and the axial-vector meson $\Dso$ have garnered significant attention because their masses are significantly below those predicted by the quark model~\cite{Godfrey:1985xj,Godfrey:1986wj,DiPierro:2001dwf} for $c\bar{s}$ mesons with their respective $\jp$ quantum numbers. Several theoretical frameworks have been proposed to explain the nature of $\Dsz$ and $\Dso$, including molecular states~\cite{Chen:2004dy,Guo:2006fu,Gamermann:2006nm,Gamermann:2007bm,Cleven:2014oka},  quark-antiquark configurations~\cite{vanBeveren:2003kd,vanBeveren:2003jv,Coito:2011qn,Hwang:2004cd,Hwang:2005tm,Simonov:2004ar,Lee:2004gt,Zhou:2011sp,Badalian:2007yr,Bardeen:2003kt,Nowak:2003ra,Kolomeitsev:2003ac,Colangelo:2003vg,Azimov:2004xk,Liu:2006jx}, tetraquark structures~\cite{Cheng:2003kg,Dmitrasinovic:2004cu,Dmitrasinovic:2012zz,Hayashigaki:2004st,Maiani:2004vq,Bracco:2005kt}, and mixed states~\cite{Browder:2003fk,Lu:2006ry,Bicudo:2005de,MartinezTorres:2011pr,Mohler:2013rwa,Tang:2023yls}. Despite these efforts, the precise nature of $\Dsz$ and $\Dso$ remains unresolved, underscoring the need for new and improved experimental data.

The $\Dsz$ has been a subject of intense experimental and theoretical interest since its discovery by the BaBar collaboration in the decay mode $D_{s}^{+}\pi^{0}$~\cite{BaBar:2003oey}, later confirmed by CLEO~\cite{CLEO:2003ggt} and Belle~\cite{Belle:2003kup}. Its mass, $2317.8 \pm 0.5$ MeV/$c^{2}$, lies below the $DK$ threshold, restricting its decay to the isospin-violating strong decay channel $\Hd$. This channel has been measured with a branching fraction of $1.00^{+0.00}_{-0.20}$ by BESIII~\cite{BESIII:2017vdm}. Radiative transitions are particularly sensitive probes of the internal structure of such hadrons, as they involve electromagnetic interactions that are well understood~\cite{Chen:2022asf}. CLEO~\cite{CLEO:2003ggt}, Belle~\cite{Belle:2003kup}, and BaBar~\cite{BaBar:2006eep} searched for $\Rd$ using 13.5\ fb$^{-1}$, 86.9 fb$^{-1}$, and 232 fb$^{-1}$ data samples, respectively, at center-of-mass (c.m.)\ energies near 10.6 GeV, but did not find any evidence for this channel. 
The most restrictive upper limit on the ratio $\BrR$ is set to 5.9\% at 90\% confidence level by CLEO with a $\Rd$ signal yield of $-6.5 \pm 5.2$~\cite{CLEO:2003ggt}. Assuming its spin-parity is $0^+$, the $D^{*}_{s0}(2317)^{+} \to D_{s}^{+} \gamma$ decay is forbidden. Though the decay width of $\Dsz$ is unknown, a determination of the branching fraction ratio $\BrR$ would provide a direct experimental constraint on various theoretical models used to explain the nature of $\Dsz$. For instance, a ratio in the range of 0.5\% to 4.25\% would strongly favor molecular interpretations~\cite{Faessler:2007gv,Fu:2021wde,Lutz:2007sk}, while a larger value ($> 8.1\%$) would align more closely with conventional $c\bar{s}$ configurations \cite{Godfrey:2003kg,Bardeen:2003kt}. Therefore, a precise measurement of this ratio is essential to resolve the nature of the $\Dsz$ meson.


In this Letter, we report the first observation of the radiative decay $\Rd$, with $\Dss \to \Ds \gamma$. The rate for this decay is measured relative to the hadronic decay $\Hd$, using 980.4~$\rm fb^{-1}$ of Belle data collected at c.m.\ energies near the $\Upsilon(nS)$ ($n = 1,...,5$) resonances, and 427.9~$\rm fb^{-1}$ of Belle II data collected at or near the c.m.\ energies of $\Upsilon(4S)$ and 10.75 GeV. Inclusion of charge conjugate states is implicit. The $D_{s}^{+}$ candidates are reconstructed via the $\phi \pi^{+}$ and $K^{+} \bar{K}^{*0}$ decay modes, both of which result in the $K^{+}K^{-}\pi^{+}$ final state. The hadronic decay serves as a reference channel, enabling cancellation of the systematic uncertainties associated with $D_{s}^{+}$ and $\gamma$ selection in the branching fraction ratio measurement.

The Belle detector~\cite{Belle:2000cnh,PTEP_belle} was a large-solid-angle spectrometer that operated at the KEKB asymmetric-energy $e^{+}e^{-}$ collider~\cite{Kurokawa:2001nw,PTEP_kekb}. The Belle~II detector~\cite{Belle-II:2010dht} is a significant upgrade of Belle and operates at the SuperKEKB $e^{+}e^{-}$ collider~\cite{Akai:2018mbz}. The $z$ axis is defined parallel to the $e^+$ beam at Belle and to the principal axis of the solenoid at Belle II with the interaction point as the origin of the coordinate system.

Data and simulated Monte Carlo (MC) samples both for Belle and Belle II are processed with the Belle~II analysis software framework~\cite{basf2,basf2_repo,Gelb:2018agf}. MC simulations are used to optimize selection criteria, investigate background sources, calculate reconstruction efficiencies, and determine the probability density functions (pdfs) employed in fitting the data. The MC events for the continuum $e^+ e^- \to c\bar{c}$ process are generated with {\sc KKMC}~\cite{Jadach:1999vf} and {\sc PYTHIA}~\cite{pythia1,pythia2}, where at least one of the charm quarks hadronizes into a $\Dsz$ meson for the signal events. The $\Rd$ and $\Hd$ decays are simulated with the phase space model, while the decay $D_{s}^{*+} \to D_{s}^{+}\gamma$ is simulated as a $P$-wave decay. The decay $D_{s}^{+} \to K^{+} K^{-} \pi^{+}$ is modeled based on previous measurements~\cite{PDG,BESIII:2020ctr}. Simulated events undergo detector simulation with \textsc{GEANT3}~\cite{geant3} for Belle and \textsc{GEANT4}~\cite{geant4} for Belle II. The signal MC samples are corrected with a reweighting method based on the measured $x_p$ distribution from the reference channel, where $x_p \equiv p^{*}_{\Dsz} / p^{*}_{\rm max}$ is the reduced momentum of the selected $\Dsz$ candidate. Here, $p^{*}_{\Dsz}$ is its momentum in the c.m.\ frame, and $p^{*}_{\rm max} \equiv \sqrt{E_{\rm beam}^{2}/c^{2} - M^{2}(\Dsz)c^{2}}$ is the maximum kinematically-allowed momentum, and $E_{\rm beam}$ is the beam energy and $M(\Dsz)$ is the invariant mass of the $\Dsz$ candidates. 

To study backgrounds, we use MC samples generated with the Belle and Belle II configurations, which correspond to four times the sizes of the corresponding datasets. Belle's MC samples include $\Upsilon(1S,2S,3S)$ decays, $\Upsilon(4S) \to B\bar{B}$, $\Upsilon(5S) \to B_{(s)}^{(*)} \bar{B}_{(s)}^{(*)}, B\bar{B}^{(*)}\pi$, and $\EE \to q\bar{q}$ $(q=u,~d,~s,~c)$ at c.m.\ energies of $\sqrt{s} = $ 10.52, 10.58, and 10.867~GeV. Belle II's MC samples consist of $\EE \to q\bar{q}$ and $\Upsilon(4S) \to B\bar{B}$.

For the signal event selection, tracks are required to satisfy $dr<0.5$~cm and $|dz|<3.0$~cm, where $dr$ and $dz$ are transverse and longitudinal impact parameters, respectively. For charged particle identification, information from different subdetectors is combined to form the likelihood $\mathcal{L}_{i}$ for species $i$, where $i=\pi$ or $K$~\cite{pid,Belle-II:2025tpe}. A track with a likelihood ratio $\mathcal{L}_K/(\mathcal{L}_K + \mathcal{L}_\pi)> 0.6~(<0.4)$ is identified as a kaon (pion). With this selection, for Belle, the kaon (pion) identification efficiency is about 88\% (90\%), while 8\% (8\%) of the pions (kaons) are misidentified as kaons (pions); for Belle II, the identification efficiency is about 97\% (98\%) for kaon (pion), with 1.3\% (1.7\%) of the pions (kaons) are misidentified as kaons (pions).

The electromagnetic calorimeter clusters not associated to tracks with energy greater than 0.10 GeV in the c.m.\ frame are regarded as photons. For the signal channel $D^{*}_{s0}(2317)^{+} \to D_{s}^{*+}(\to D_{s}^{+}\gamma_{2}) \gamma_{1}$, the energy of $\gamma_{1}$ is required to be greater than 0.22 GeV in the c.m. frame. Additionally, the invariant mass $M_{\gamma_{1}\gamma_{2}}$ must lie outside the region $[0.10,~0.16]$ GeV/$c^2$ to exclude $\Hd$ candidates. For the reference channel $\Hd$, one of the signal photons should have energy greater than 0.22 GeV in the c.m.\ frame, and $M(\gamma_1\gamma_2)$ must be within 15 MeV/$c^{2}$ of the known $\pi^{0}$ mass to form $\pi^0$ candidates ($\sim 2.5\sigma$)~\cite{PDG}. For both decay channels, the invariant mass of any combination of a signal photon and any other photon in the event must not fall within 15 MeV/$c^{2}$ of the known $\pi^{0}$ mass~\cite{PDG}. 
    
The $K^{+}$, $K^{-}$, and $\pi^{+}$ candidates are combined to form $D_{s}^{+}$ candidates. For the decay channel $D_{s}^{+} \to \phi \pi^{+}$, we require the invariant mass of the $K^{+}K^{-}$ pair to satisfy $|M(K^{+}K^{-}) - m(\phi)| < 0.01$ GeV/$c^{2}$ ($\sim 2.5\sigma$), where $m(\phi)$ is the known $\phi$ mass~\cite{PDG}. For the decay channel $D_{s}^{+} \to K^{+} \bar{K}^{*0}$, the invariant mass of the $K^{-}\pi^{+}$ pair must satisfy $|M(K^{-}\pi^{+}) - m(K^{*0})| < 0.05$ GeV/$c^{2}$, where $m(K^{*0})$ is the known $K^{*0}$ mass~\cite{PDG}. The invariant mass of the $K^{+}K^{-}\pi^{+}$ system must satisfy $|M(K^{+}K^{-}\pi^{+}) - m(D_{s}^{+})| < 0.01$ GeV/$c^2$,  where $m(D_{s}^{+})$ is the known $D_{s}^{+}$ mass~\cite{PDG}. 

The $\Ds$ candidates are combined with a photon to form $\Dss$ candidates. The mass window of $\Dss$ is $|M(\Ds\gamma_2) - m(\Dss) | <$ 0.015 GeV/$c^{2}$, where $m(\Dss)$ is the known mass of $\Dss$~\cite{PDG}, and $M(\Ds\gamma_{2}) = M^{\rm rec}(\Ds\gamma_2) - M^{\rm rec}(K^{+}K^{-}\pi^{+}) + m(\Ds)$ is used to cancel the contribution to the mass resolution from the measurement of $\Ds$. Here and below, we use $M^{\rm rec}(X)$ to indicate the reconstructed invariant mass of the $X$ system. After applying these requirements, there are no candidates for which $M(\Ds\gamma_1)$ falls within the $\Dss$ mass window. Then, the combinations of $\Dss\gamma_{1}$ or $\Hdd$ are considered as $\Dsz$ candidates. 

\begin{figure}[htbp]
    \begin{center}
    \includegraphics[width=0.4\textwidth]{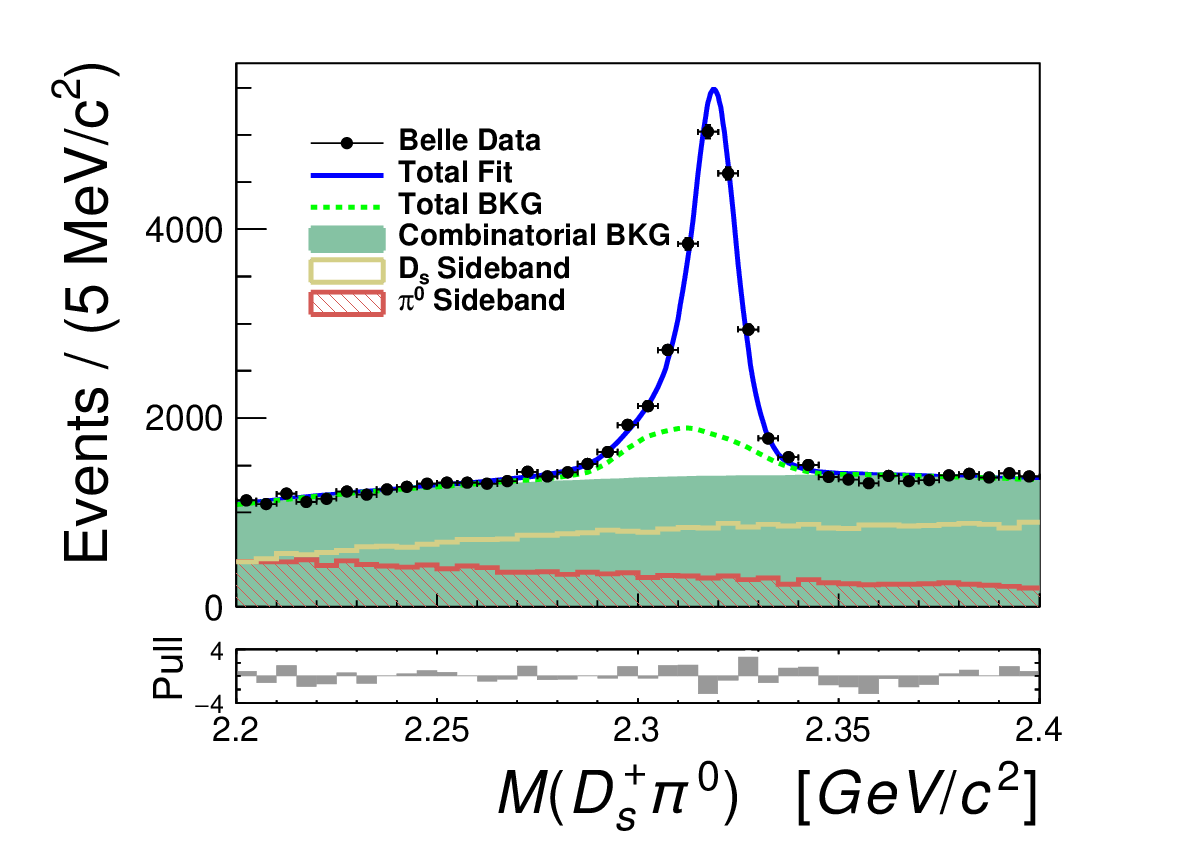}
    \put(-60,110){\Large (a)}
    
    \includegraphics[width=0.4\textwidth]{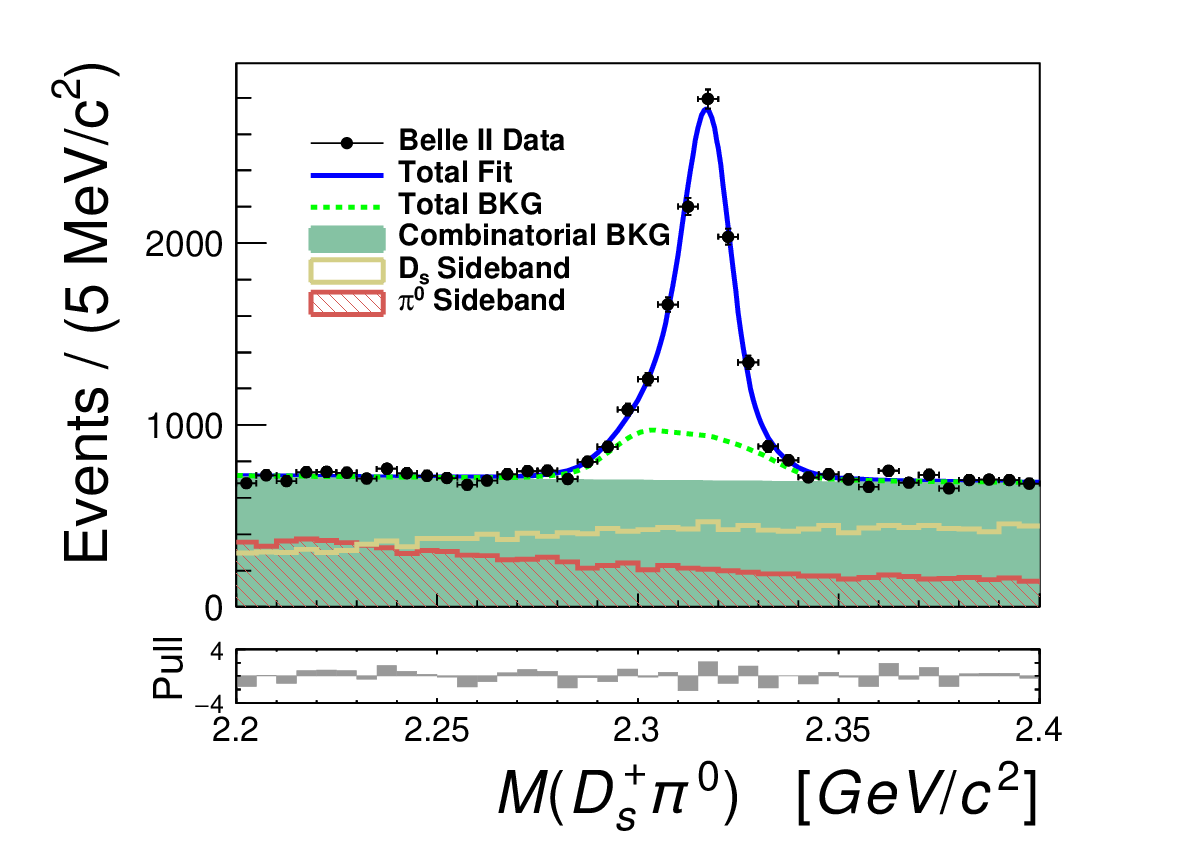}
    \put(-60,110){ \Large (b)} 
    \caption{ Invariant mass distributions of $\Hdd$ from (a) Belle and (b) Belle II data. The data samples are represented by the dots with error bars. The blue curves, green dotted curves, and green filled areas are the fitted total pdfs, total backgrounds, and combinatorial backgrounds, respectively. The areas between total and combinatorial backgrounds are from the fitted cross-feed contributions. The distributions from the normalized $\Ds$ and $\piz$ mass sidebands are shown with yellow blank and red slash filled histograms, respectively.} \label{Mhd_Fit}
    \end{center}
\end{figure}

\begin{figure}[htbp]
    \begin{center}
    \includegraphics[width=0.4\textwidth]{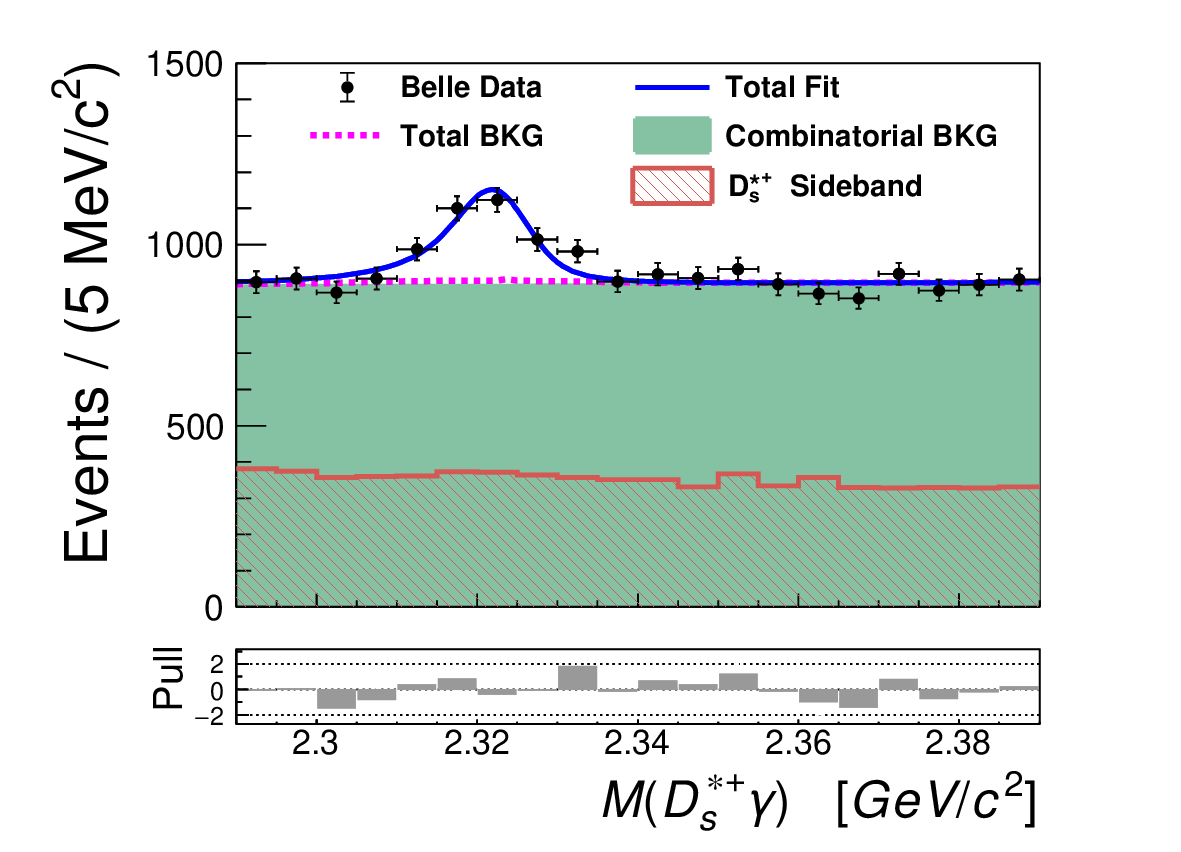}
    \put(-155,110){ (a)}
    
    \includegraphics[width=0.4\textwidth]{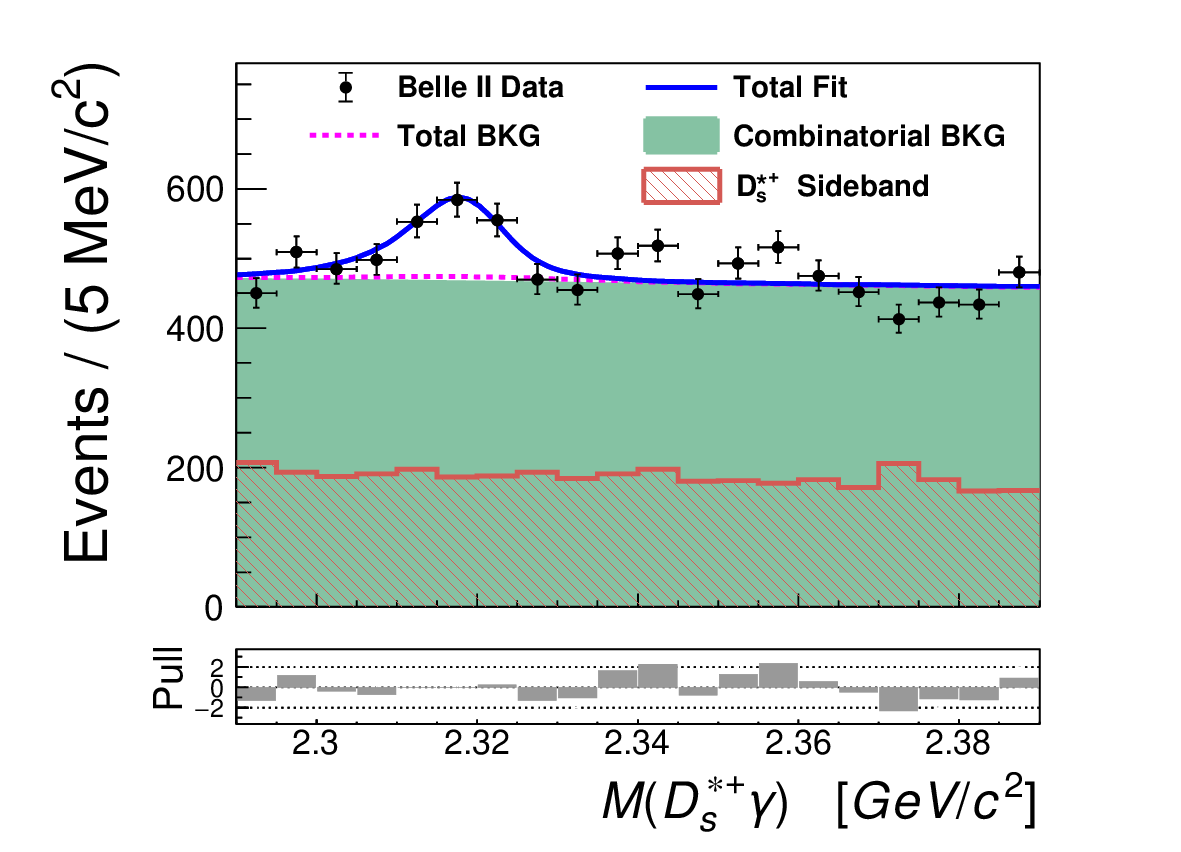}
    \put(-155,110){ (b)} 
  \caption{Simultaneous fits to the $M(\Dss\gamma)$ distributions from (a) Belle and (b) Belle II. The blue and violet curves are the best fit results and the fitted total background pdfs, respectively. The filled green areas are the fitted combinatorial backgrounds. The histograms in red slashes represent the normalized $\Dss$ sidebands.} \label{Mrd_Fit}
    \end{center} 
\end{figure}

To suppress the combinatorial background, we require $x_p$ to be larger than 0.7, which also removes all $\Dsz$ from $B$ decays. The $D_{s}^{+}$ and $D_{s}^{*+}$ signal purities are  81\% (86\%) and 68\% (63\%) for Belle (Belle II) for the radiative decay channel, respectively. All the possible candidates in an event are retained for further analysis, with the multiplicity of 1.03 (1.02) for radiative (hadronic) decay channel in the studied regions shown in Fig.~\ref{Mhd_Fit} and~\ref{Mrd_Fit}.

We optimize the selection criteria by maximizing the Punzi figure of merit $\varepsilon/(5/2 + \sqrt{N_{B}})$~\cite{Punzi:2003bu} in the signal region ($2.29 < M(\Dss\gamma) < 2.34$~GeV/$c^{2}$) of the $\Rd$ channel, where $\varepsilon$ is the detection efficiency. The background yield $N_{B}$ is estimated in a data-driven way by linearly extrapolating the yield from the upper sideband ($2.35 < M(\Dss\gamma) < 2.40$~GeV/$c^{2}$), as the lower sideband contains a peaking background due to a random photon combining with a real $\Ds$ candidate to form the fake $\Dss$ candidate, while the true $\gamma_{2}$ is mistakenly selected as $\gamma_{1}$. Similar peaking backgrounds have been reported previously by the CLEO experiment~\cite{CLEO:2003ggt}. Since modeling the shapes of these backgrounds is challenging and they do not affect the extracted signal yields of the $\Dsz$, we exclude this lower mass region in this analysis. 

Here, we use $M(\Dss\gamma) = M^{\rm rec}(\Dss\gamma_{1}) - M^{\rm rec}(\Ds\gamma_{2}) + m(\Dss)$ as this cancels the contribution to the mass resolution from the measurement of the $\Dss$. Following a blind analysis strategy, we do not examine the $M(\Rdd)$ distributions in the signal region until the analysis procedure is finalized.

After applying the aforementioned selections, the $M(\Hdd)$ distributions are presented in Fig.~\ref{Mhd_Fit}, revealing distinct peaks corresponding to the $\Dsz$ state in both datasets. Here, we use $M(\Hdd) = M^{\rm rec}(\Hdd) - M^{\rm rec}(K^{+}K^{-}\pi^{+}) + m(\Ds) - M^{\rm rec}(\gamma_{1}\gamma_{2}) + m(\pi^0)$ as this cancels the contributions to the mass resolutions from the
measurements of the $\Ds$ and $\pi^0$. The {\sc TopoAna} package~\cite{zhouxy_topo} is used for the backgrond study. Based on the study of the MC samples, apart from the combinatorial background, the $\Dso \to \Dss \piz$ decay with the photon from $\Dss$ missed can introduce an excess under the $\Dsz$ peak (denoted as the cross-feed). The distributions from the normalized $\Ds$ and $\piz$ mass sidebands ( $|M(K^{+}K^{-}\pi^{+}) - m(\Ds) \pm 0.04| < 0.01$ GeV/$c^{2}$ and $|M(\gamma\gamma) - m(\pi^{0}) \pm 0.05| < 0.0075$ GeV/$c^{2}$) are shown with yellow blank and red slash filled histograms in Fig. \ref{Mhd_Fit}, which exhibit no peaking structures, i.e. the background from the $\Rd$ channel is negligible. 

The signal yields of $\Hd$ are extracted from the unbinned extended maximum-likelihood fits to the $M(\Hdd)$ spectra. In each fit, the signal pdf is represented by a Crystal Ball (CB) function~\cite{CB_fun} convolved with a triple-Gaussian function, whose parameters, except the mean values of the CB functions, are fixed according to signal MC simulations. The cross-feed pdf is constructed from smoothed histograms of MC events. The combinatorial backgrounds are described by the second-order Chebyshev polynomials. The yields of these components are floated in the fits, and the fit results are shown in Fig.~\ref{Mhd_Fit}. The fit method is validated by the MC samples. The similar fits are performed to $M(\Hdd)$ spectra from different $x_p$ bins to measure the $x_p$ distribution of $\Dsz$. The obtained efficiency-corrected $x_p$ distribution is used to correct the MC simulation. The fitted yields of the hadronic decay channel $N^{\rm fit}_{\rm exp}(\Hdd)$ are $10820 \pm 230$ for Belle and $6108 \pm 163$ for Belle II. For events with $x_p > 0.7$, the detection efficiencies $\varepsilon_{\rm exp}(\Hdd)$ are 4.6\% and 5.2\% for Belle and Belle II, respectively. Here and after, the subscript ${\rm exp}$ indicates an experiment (Belle or Belle II).


For the $\Rd$ channel, the $M(\Rdd)$ spectra from Belle and Belle II data are presented in Fig.~\ref{Mrd_Fit}, where the $\Dsz$ signal peak is clearly visible in both plots. According to the studies done on MC simulations~\cite{zhouxy_topo}, we don’t anticipate any peaking contribution from $\Hd$ as well as from $\Dso \to \Dss\pi^0,\Dss\gamma,$ and $\Dsz\gamma$ decays. Furthermore, no peaking in the $\Dsz$ signal region is found in the normalized $\Dss$ mass sidebands ($|M(\Ds\gamma) - m(\Dss)$ $ \pm 0.05 | <$ 0.015 GeV/$c^{2}$). A small peaking background, which we label as `broken signal' , may arise when a correctly reconstructed $\Ds \gamma_{1}$ candidate is combined with a wrongly selected $\gamma_{2}$. 

 We extract the branching fraction ratio $\BrR$, denoted ${\cal R}$, through a simultaneous unbinned extended maximum-likelihood fit to the $M(\Rdd)$ spectra from Belle and Belle II, as shown in Fig.~\ref{Mrd_Fit}. Each $\Dsz$ signal pdf is modeled by a CB function convolved with a triple-Gaussian function, while the corresponding broken signal contribution is described by an asymmetric Gaussian. All parameters of the broken signal and signal pdf, as well as the ratios of their yields, are fixed from MC simulations, except for the mean values of the CB functions. The yield ratio of the broken signal to signal component is 7.5\% (9.3\%) for Belle (Belle II). The combinatorial backgrounds are described by 1st-order polynomials. The mass resolution is approximately 7 MeV for both Belle and Belle II. The value of $\mathcal{R}$ is shared as a common free parameter in the simultaneous fit, while the $\Rd$ signal yields are set according to $N_{\rm exp}(\Rdd) = {\cal R} N^{\rm fit}_{\rm exp}(\Hdd) \varepsilon_{\rm exp}(\Rdd)/\varepsilon_{\rm exp}(\Hdd)$ separately for Belle and Belle II. 
Here, $\varepsilon_{\rm exp}(\Rdd)$ is the detection efficiency of $\Rd$ decay, which is 4.2\% for Belle and 4.6\% for Belle II. The fit results are shown in Fig.~\ref{Mrd_Fit}. The fitted masses of the
$\Dsz$ in the Belle and Belle II datasets differ by $3.6\pm1.5$ MeV/$c^2$. 
This mass difference is consistent with known discrepancies in low-energy photon calibration between the two detectors and does not affect the  ${\cal R}$ measurement, as the simultaneous fit accounts for this offset through independent mean values for each dataset. The fitted ${\cal R}$ value is $[7.13 \pm 0.70 ({\rm stat.})]\%$. The corresponding $N_{\rm exp}(\Rdd)$ are $712\pm 69$ and $387 \pm 38$ for Belle and Belle II, respectively. 

The significance of $\Rd$ is 10.1$\sigma$, estimated from the negative log-likelihood ratio $-2\ln(\mathcal{L}_0/ \mathcal{L}_{\rm max}) = 111.9$~\cite{significance} with the difference in degrees of freedom ($\Delta{\rm d.o.f.} = 3$) and the systematic uncertainty discussed below considered. Here, $\mathcal{L}_0$ and $\mathcal{L}_{\rm max}$ represent the maximized likelihoods of the simultaneous fits without and with the $\Rd$ signal components, respectively. Systematic uncertainties are incorporated by convolving the likelihood ratio distribution with a Gaussian function whose width corresponds to the total systematic uncertainty. We also perform separate fits to the Belle and Belle II data using the same fit components as those in the simultaneous fit. The fitted signal yields $N^{\rm fit}_{\rm exp}(\Rdd)$ are $742\pm82$ and $348\pm69$ for Belle and Belle II, respectively. 
The corresponding ${\cal R}$ values are $[7.43 \pm 0.83 ({\rm stat.}) ]\%$ and $[6.43 \pm 1.27 ({\rm stat.})]\%$, demonstrating good consistency between the results of the simultaneous fit and the fits to each individual dataset.

The systematic uncertainties due to $D_s^+$ and $\gamma$ selection cancel in the ${\cal R}$ measurement. The dominant systematic uncertainties are from the fit model and $x_p$ weighting. All systematic sources are described below and the resulting percent variations relative to the nominal fit are listed in Table \ref{Tab:BR_syst}. 

To characterize possible systematic effects in the $\Hd$ reference channel, (a) the order of the background polynomials and the fit range are varied, (b) the widths of the triple-Gaussian functions are increased by 1$\sigma$, and (c) the resolution pdf variations are also propagated into the modeling of the $\Dso \to \Dss\piz$ cross-feed components. Finally, the differences in the fitted $\Hd$ yields are taken as systematic uncertainties of $N^{\rm fit}_{\rm exp}(\Hdd)$, which are 1.3\% (1.0\%),  0.7\% (1.5\%), and 0.8\% (0.3\%), from the fit region/background pdf, resolution, and cross-feed pdf for Belle (Belle II), respectively. 

A series of pseudo-experiments is conducted to estimate the systematic uncertainty contribution to ${\cal R}$ from the $\Hd$ channel. In each trial, we randomly fluctuate the $\Hd$ yields for both Belle and Belle II by sampling from Gaussian distributions. Each Gaussian distribution is constructed with its mean value set to the corresponding nominal $\Hd$ yield and its standard deviation equal to the systematic uncertainty of $N^{\rm fit}_{\rm exp}(\Hdd)$. 
A simultaneous fit similar to the nominal fit to the data described above is then performed to the $M(\Rdd)$ distributions from data for each set of the pseudo-yields of $\Hd$. From these results, an ensemble of Gaussian-distributed varied ${\cal R}$ values is obtained whose width is taken as the systematic uncertainty on ${\cal R}$ from $\Hd$ decay.

We characterize systematic effects in the signal channel $\Rd$ fits by examining the changes of fitted ${\cal R}$ values in the simultaneous fit to $M(\Rdd)$ distributions from data after (a) varying the order of the background polynomials and the fit range, (b) increasing the widths of the triple-Gaussian functions by 1$\sigma$, and (c) adjusting the ratios and widths of broken signal to signal yields by $2\sigma$ to conservatively estimate the systematic uncertainty. The differences of the fitted $\cal{R}$ values from the nominal result are taken as systematic uncertainties. 

To estimate the uncertainty due to $x_{p}$ reweighting, we vary the polynomial order when fitting the efficiency-corrected $x_{p}$ distribution and reweight the signal MC samples accordingly. Then, the new detection efficiencies derived from the reweighted signal MC samples are used in the simultaneous fit to $M(\Rdd)$ distributions from data. The change of the fitted $\cal{R}$ value from the nominal result is taken as the systematic uncertainty.

By reweighting the $\cos\theta_{\gamma_{1}\gamma_{2}}$ distribution in the $\Dsz$ signal MC samples to match a $1+\cos^{2}\theta_{\gamma_{1}\gamma_{2}}$ distribution, we obtained a new set of detection efficiencies. Here, $\theta_{\gamma_{1}\gamma_{2}}$ is the angle between the two photons in the $D_{s}^{*}$ rest frame. The corresponding systematic uncertainty is estimated using a method similar to that for the $x_{p}$ reweighting.

The systematic uncertainty on detection efficiencies due to the limited size of the signal MC sample is estimated by $\sqrt{(1-\varepsilon)\varepsilon/N}$, where $\varepsilon$ and $N$ are the detection efficiency and number of simulated signal events, respectively. By varying the detection efficiencies by $1\sigma$ in the simultaneous fit to $M(\Rdd)$ from data, the change of the fitted $\cal{R}$ from the nominal result is taken as the systematic uncertainty.

Assuming that all the systematic uncertainties detailed above are independent,
they are added in quadrature to obtain the total systematic uncertainty of 3.2\%, as listed in Table~\ref{Tab:BR_syst}.

\begin{table}[htbp]
	\caption{\label{Tab:BR_syst} The summary of the systematic uncertainties of the measurement of the branching fraction ratio $\BrR$ (in \%).}

	\begin{center}
	\tabcolsep=200pt
	\setlength{\tabcolsep}{10pt}
		\begin{tabular}{lcc}
			\toprule[1pt]
			 Source &  $\Ds \piz$&  $\Dss\gamma$   \\
			 \hline
			 Fit region and background pdf & 0.8 & 1.3\\
			 Fixed pdf parameters & 0.7 & 2.5\\
			 Cross-feed or broken signal & 0.6 &  0.7\\
			 $x_{p}$ reweighting & \multicolumn{2}{c}{0.5 }\\
			 $\cos\theta_{\gamma_{1}\gamma_{2}}$ reweighting & \multicolumn{2}{c}{1.7 }\\
			 MC sample size & \multicolumn{2}{c}{0.5}\\
			 \hline
			 Sum & \multicolumn{2}{c}{3.6} \\
			\toprule[1pt]
		\end{tabular}	    
	
\end{center}

\end{table}		
 
 \begin{figure}[htbp]
 	\begin{minipage}{0.47\textwidth}
 		\centerline{\includegraphics[width=1\linewidth]{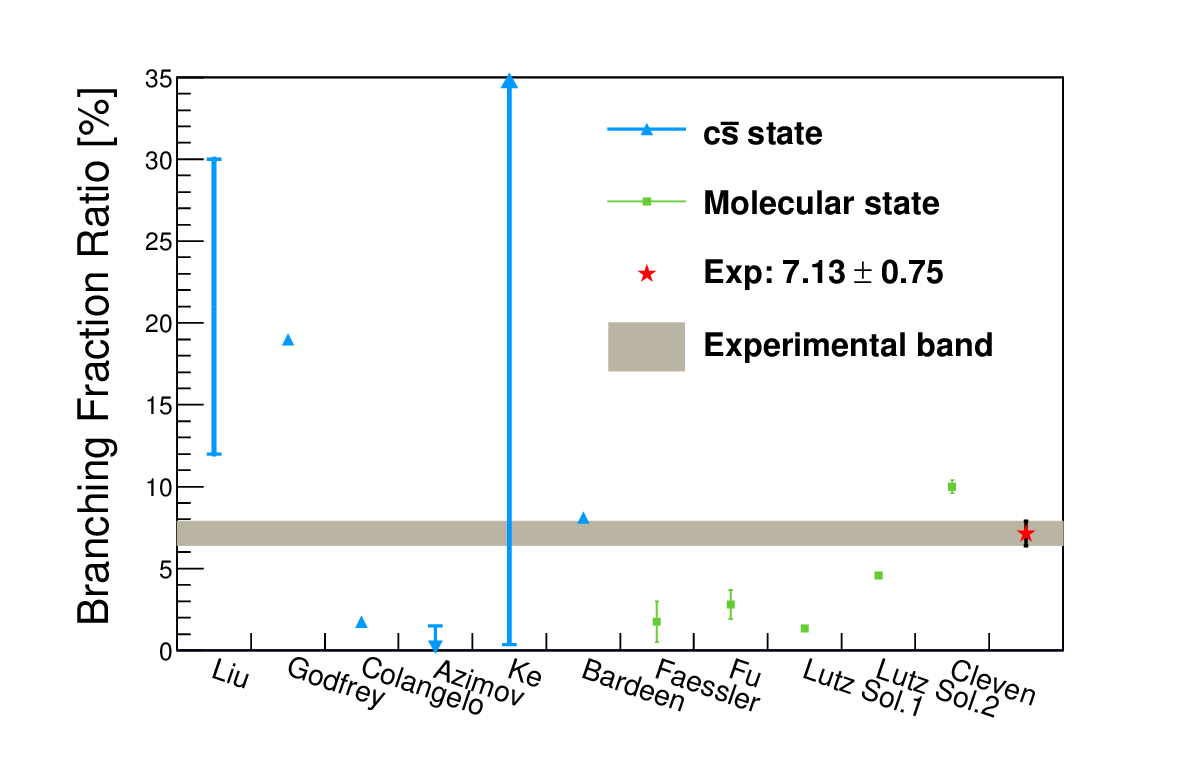}}	
 	\end{minipage}
 	\caption{Comparison of the measured $\BrR$ with theoretical predictions. The theoretical approaches of the references are effective field theory for Liu~\cite{Liu:2006jx} and Cleven~\cite{Cleven:2014oka}, traditional quark model for Godfrey~\cite{Godfrey:2003kg}, heavy quark symmetries for Colangelo~\cite{Colangelo:2003vg}, phenomenological for Azimov~\cite{Azimov:2004xk}., light front quark model for Ke~\cite{Ke:2013zs}, effective Lagrangian with with chiral symmetry for Bardeen~\cite{Bardeen:2003kt}, effective Lagrangian for Faessler~\cite{Faessler:2007gv}, heavy quark flavor symmetry for Fu~\cite{Fu:2021wde},  and chiral lagrangian with coupled-channel dynamics for Lutz~\cite{Lutz:2007sk}. The prediction of Ke~\cite{Ke:2013zs} is a lower limit indicated by an arrow. The uncertainty of the experimental measurement is the quadrature sum of statistical and systematic uncertainties.
 	} \label{Fig:Sum}
 \end{figure}

In summary, using combined data samples of 980~$\rm fb^{-1}$ from Belle and 428~$\rm fb^{-1}$ from Belle~II collected at the KEKB and SuperKEKB asymmetric-energy $\EE$ colliders, respectively, we have made the first observation of the radiative decay $\Rd$ in the continuum $e^+ e^-$ $ \to$ $c\bar{c}$ process with a significance exceeding 10 standard deviations. A comparison between theoretical predictions and the measured $\BrR$ value is presented in Fig.~\ref{Fig:Sum}. The branching fraction ratio $\BrR$ is measured to be $[7.13 \pm 0.70({\rm stat.}) \pm 0.26({\rm syst.})]\%$, which is generally larger than theoretical predictions suggesting $\Dsz$ as a molecular state~\cite{Faessler:2007gv,Fu:2021wde,Lutz:2007sk}, while smaller than the $c\bar{s}$ state assumption under the quark model~\cite{Godfrey:2003kg}, and under effective field theory for both assumption~\cite{Liu:2006jx,Cleven:2014oka} . However, predictions based on the light front quark model~\cite{Ke:2013zs} and chiral quark model~\cite{Bardeen:2003kt}  agree with our measurement under the pure $c\bar{s}$ state expectation. While most theoretical models address an anomalously low mass of the $\Dsz$ and provide estimates of its width, the intrinsic width cannot currently be measured due to limited experimental resolution. Our result provides a complementary input that will help to further discriminate among competing explanations for the nature of the $\Dsz$ and motivates dedicated theoretical calculations of the ratio $\BrR$ under various hypotheses.


This work, based on data collected using the Belle II detector, which was built and commissioned prior to March 2019, and data collected using the Belle detector, which was operated until June 2010, was supported by
Higher Education and Science Committee of the Republic of Armenia Grant No.~23LCG-1C011;
Australian Research Council and Research Grants
No.~DP200101792, 
No.~DP210101900, 
No.~DP210102831, 
No.~DE220100462, 
No.~LE210100098, 
and
No.~LE230100085; 
Austrian Federal Ministry of Education, Science and Research,
Austrian Science Fund (FWF) Grants
DOI:~10.55776/P34529,
DOI:~10.55776/J4731,
DOI:~10.55776/J4625,
DOI:~10.55776/M3153,
and
DOI:~10.55776/PAT1836324,
and
Horizon 2020 ERC Starting Grant No.~947006 ``InterLeptons'';
Natural Sciences and Engineering Research Council of Canada, Digital Research Alliance of Canada, and Canada Foundation for Innovation;
National Key R\&D Program of China under Contract No.~2024YFA1610503,
and
No.~2024YFA1610504
National Natural Science Foundation of China and Research Grants
No.~11575017,
No.~11761141009,
No.~11705209,
No.~11975076,
No.~12135005,
No.~12150004,
No.~12161141008,
No.~12405099,
No.~12475093,
No.~12175041,
and
No. 12405102.
and Shandong Provincial Natural Science Foundation Project~ZR2022JQ02;
the Czech Science Foundation Grant No. 22-18469S,  Regional funds of EU/MEYS: OPJAK
FORTE CZ.02.01.01/00/22\_008/0004632 
and
Charles University Grant Agency project No. 246122;
European Research Council, Seventh Framework PIEF-GA-2013-622527,
Horizon 2020 ERC-Advanced Grants No.~267104 and No.~884719,
Horizon 2020 ERC-Consolidator Grant No.~819127,
Horizon 2020 Marie Sklodowska-Curie Grant Agreement No.~700525 ``NIOBE''
and
No.~101026516,
and
Horizon 2020 Marie Sklodowska-Curie RISE project JENNIFER2 Grant Agreement No.~822070 (European grants);
L'Institut National de Physique Nucl\'{e}aire et de Physique des Particules (IN2P3) du CNRS
and
L'Agence Nationale de la Recherche (ANR) under Grant No.~ANR-23-CE31-0018 (France);
BMFTR, DFG, HGF, MPG, and AvH Foundation (Germany);
Department of Atomic Energy under Project Identification No.~RTI 4002,
Department of Science and Technology,
and
UPES SEED funding programs
No.~UPES/R\&D-SEED-INFRA/17052023/01 and
No.~UPES/R\&D-SOE/20062022/06 (India);
Israel Science Foundation Grant No.~2476/17,
U.S.-Israel Binational Science Foundation Grant No.~2016113, and
Israel Ministry of Science Grant No.~3-16543;
Istituto Nazionale di Fisica Nucleare and the Research Grants BELLE2,
and
the ICSC – Centro Nazionale di Ricerca in High Performance Computing, Big Data and Quantum Computing, funded by European Union – NextGenerationEU;
Japan Society for the Promotion of Science, Grant-in-Aid for Scientific Research Grants
No.~16H03968,
No.~16H03993,
No.~16H06492,
No.~16K05323,
No.~17H01133,
No.~17H05405,
No.~18K03621,
No.~18H03710,
No.~18H05226,
No.~19H00682, 
No.~20H05850,
No.~20H05858,
No.~22H00144,
No.~22K14056,
No.~22K21347,
No.~23H05433,
No.~26220706,
and
No.~26400255,
and
the Ministry of Education, Culture, Sports, Science, and Technology (MEXT) of Japan;  
National Research Foundation (NRF) of Korea Grants
No.~2021R1-F1A-1064008, 
No.~2022R1-A2C-1003993,
No.~2022R1-A2C-1092335,
No.~RS-2016-NR017151,
No.~RS-2018-NR031074,
No.~RS-2021-NR060129,
No.~RS-2023-00208693,
No.~RS-2024-00354342
and
No.~RS-2025-02219521,
Radiation Science Research Institute,
Foreign Large-Size Research Facility Application Supporting project,
the Global Science Experimental Data Hub Center, the Korea Institute of Science and
Technology Information (K25L2M2C3 ) 
and
KREONET/GLORIAD;
Universiti Malaya RU grant, Akademi Sains Malaysia, and Ministry of Education Malaysia;
Frontiers of Science Program Contracts
No.~FOINS-296,
No.~CB-221329,
No.~CB-236394,
No.~CB-254409,
and
No.~CB-180023, and SEP-CINVESTAV Research Grant No.~237 (Mexico);
the Polish Ministry of Science and Higher Education and the National Science Center;
the Ministry of Science and Higher Education of the Russian Federation
and
the HSE University Basic Research Program, Moscow;
University of Tabuk Research Grants
No.~S-0256-1438 and No.~S-0280-1439 (Saudi Arabia), and
Researchers Supporting Project number (RSPD2025R873), King Saud University, Riyadh,
Saudi Arabia;
Slovenian Research Agency and Research Grants
No.~J1-50010
and
No.~P1-0135;
Ikerbasque, Basque Foundation for Science,
State Agency for Research of the Spanish Ministry of Science and Innovation through Grant No. PID2022-136510NB-C33, Spain,
Agencia Estatal de Investigacion, Spain
Grant No.~RYC2020-029875-I
and
Generalitat Valenciana, Spain
Grant No.~CIDEGENT/2018/020;
The Knut and Alice Wallenberg Foundation (Sweden), Contracts No.~2021.0174, No.~2021.0299, and No.~2023.0315;
National Science and Technology Council,
and
Ministry of Education (Taiwan);
Thailand Center of Excellence in Physics;
TUBITAK ULAKBIM (Turkey);
National Research Foundation of Ukraine, Project No.~2020.02/0257,
and
Ministry of Education and Science of Ukraine;
the U.S. National Science Foundation and Research Grants
No.~PHY-1913789 
and
No.~PHY-2111604, 
and the U.S. Department of Energy and Research Awards
No.~DE-AC06-76RLO1830, 
No.~DE-SC0007983, 
No.~DE-SC0009824, 
No.~DE-SC0009973, 
No.~DE-SC0010007, 
No.~DE-SC0010073, 
No.~DE-SC0010118, 
No.~DE-SC0010504, 
No.~DE-SC0011784, 
No.~DE-SC0012704, 
No.~DE-SC0019230, 
No.~DE-SC0021274, 
No.~DE-SC0021616, 
No.~DE-SC0022350, 
No.~DE-SC0023470; 
and
the Vietnam Academy of Science and Technology (VAST) under Grants
No.~NVCC.05.02/25-25
and
No.~DL0000.05/26-27.

These acknowledgements are not to be interpreted as an endorsement of any statement made
by any of our institutes, funding agencies, governments, or their representatives.

We thank the SuperKEKB team for delivering high-luminosity collisions;
the KEK cryogenics group for the efficient operation of the detector solenoid magnet and IBBelle on site;
the KEK Computer Research Center for on-site computing support; the NII for SINET6 network support;
and the raw-data centers hosted by BNL, DESY, GridKa, IN2P3, INFN, PNNL/EMSL, 
and the University of Victoria.

\end{document}

%% file: pub100-orcid.tex
  \author{M.~Abumusabh\,\orcidlink{0009-0004-1031-5425}} 
  \author{I.~Adachi\,\orcidlink{0000-0003-2287-0173}} 
  \author{L.~Aggarwal\,\orcidlink{0000-0002-0909-7537}} 
  \author{H.~Ahmed\,\orcidlink{0000-0003-3976-7498}} 
  \author{Y.~Ahn\,\orcidlink{0000-0001-6820-0576}} 
  \author{H.~Aihara\,\orcidlink{0000-0002-1907-5964}} 
  \author{N.~Akopov\,\orcidlink{0000-0002-4425-2096}} 
  \author{S.~Alghamdi\,\orcidlink{0000-0001-7609-112X}} 
  \author{M.~Alhakami\,\orcidlink{0000-0002-2234-8628}} 
  \author{A.~Aloisio\,\orcidlink{0000-0002-3883-6693}} 
% ...
\collaboration{The Belle and Belle II Collaboration}

%% file: draft.bbl
\begin{thebibliography}{99}
\bibitem{Brambilla:2019esw} N.~Brambilla, S.~Eidelman, C.~Hanhart, A.~Nefediev, C.~P.~Shen, C.~E.~Thomas, A.~Vairo, and C.~Z.~Yuan, 
Phys. Rept. \textbf{873}, 1 (2020).

\bibitem{Chen:2016spr}
H.~X.~Chen, W.~Chen, X.~Liu, Y.~R.~Liu, and S.~L.~Zhu,
Rept. Prog. Phys. \textbf{80}, 076201 (2017).

\bibitem{Guo:2017jvc}
F.~K.~Guo, C.~Hanhart, U.~G.~Mei\ss{}ner, Q.~Wang, Q.~Zhao, and B.~S.~Zou,
Rev. Mod. Phys. \textbf{90}, 015004 (2018)
[erratum: Rev. Mod. Phys. \textbf{94}, 029901 (2022)].

\bibitem{Godfrey:1985xj}
S.~Godfrey and N.~Isgur,
Phys. Rev. D \textbf{32}, 189 (1985).

\bibitem{Godfrey:1986wj}
S.~Godfrey and R.~Kokoski,
Phys. Rev. D \textbf{43}, 1679 (1991).

\bibitem{DiPierro:2001dwf}
M.~Di Pierro and E.~Eichten,
Phys. Rev. D \textbf{64}, 114004 (2001).

\bibitem{Chen:2004dy}
Y.~Q.~Chen and X.~Q.~Li,
Phys. Rev. Lett. \textbf{93}, 232001 (2004).

\bibitem{Guo:2006fu}
F.~K.~Guo, P.~N.~Shen, H.~C.~Chiang, R.~G.~Ping, and B.~S.~Zou,
Phys. Lett. B \textbf{641}, 278 (2006).

\bibitem{Gamermann:2006nm}
D.~Gamermann, E.~Oset, D.~Strottman, and M.~J.~Vicente Vacas,
Phys. Rev. D \textbf{76}, 074016 (2007).

\bibitem{Gamermann:2007bm}
D.~Gamermann, L.~R.~Dai, and E.~Oset,
Phys. Rev. C \textbf{76}, 055205 (2007).

\bibitem{Cleven:2014oka}
M.~Cleven, H.~W.~Grie{\ss}hammer, F.~K.~Guo, C.~Hanhart and U.~G.~Mei{\ss}ner,
Eur. Phys. J. A \textbf{50}, 149 (2014).

\bibitem{vanBeveren:2003kd}
E.~van Beveren and G.~Rupp,
Phys. Rev. Lett. \textbf{91}, 012003 (2003).

\bibitem{vanBeveren:2003jv}
E.~van Beveren and G.~Rupp,
Eur. Phys. J. C \textbf{32}, 493 (2004).

\bibitem{Coito:2011qn}
S.~Coito, G.~Rupp, and E.~van Beveren,
Phys. Rev. D \textbf{84}, 094020 (2011).

\bibitem{Hwang:2004cd}
D.~S.~Hwang and D.~W.~Kim,
Phys. Lett. B \textbf{601}, 137 (2004)

\bibitem{Hwang:2005tm}
D.~S.~Hwang and D.~W.~Kim,
J. Phys. Conf. Ser. \textbf{9}, 63 (2005).

\bibitem{Simonov:2004ar}
Y.~A.~Simonov and J.~A.~Tjon,
Phys. Rev. D \textbf{70}, 114013 (2004).

\bibitem{Lee:2004gt}
I.~W.~Lee, T.~Lee, D.~P.~Min, and B.~Y.~Park,
Eur. Phys. J. C \textbf{49}, 737 (2007).

\bibitem{Zhou:2011sp}
Z.~Y.~Zhou and Z.~Xiao,
Phys. Rev. D \textbf{84}, 034023 (2011).

\bibitem{Badalian:2007yr}
A.~M.~Badalian, Y.~A.~Simonov, and M.~A.~Trusov,
Phys. Rev. D \textbf{77}, 074017 (2008).

\bibitem{Bardeen:2003kt}
W.~A.~Bardeen, E.~J.~Eichten, and C.~T.~Hill,
Phys. Rev. D \textbf{68}, 054024 (2003).

\bibitem{Nowak:2003ra}
M.~A.~Nowak, M.~Rho, and I.~Zahed,
Acta Phys. Polon. B \textbf{35}, 2377 (2004).

\bibitem{Kolomeitsev:2003ac}
E.~E.~Kolomeitsev and M.~F.~M.~Lutz,
Phys. Lett. B \textbf{582}, 39 (2004).

\bibitem{Colangelo:2003vg}
P.~Colangelo and F.~De Fazio,
Phys. Lett. B \textbf{570}, 180 (2003).

\bibitem{Azimov:2004xk}
Y.~I.~Azimov and K.~Goeke,
Eur. Phys. J. A \textbf{21}, 501 (2004).

\bibitem{Liu:2006jx}
X.~Liu, Y.~M.~Yu, S.~M.~Zhao and X.~Q.~Li,
Eur. Phys. J. C \textbf{47}, 445-452 (2006).


\bibitem{Cheng:2003kg}
H.~Y.~Cheng and W.~S.~Hou,
Phys. Lett. B \textbf{566}, 193 (2003).

\bibitem{Dmitrasinovic:2004cu}
V.~Dmitrasinovic,
Phys. Rev. D \textbf{70}, 096011 (2004).

\bibitem{Dmitrasinovic:2012zz}
V.~Dmitrasinovic,
Phys. Rev. D \textbf{86}, 016006 (2012).


\bibitem{Hayashigaki:2004st}
A.~Hayashigaki and K.~Terasaki,
Prog. Theor. Phys. \textbf{114}, 1191 (2006).

\bibitem{Maiani:2004vq}
L.~Maiani, F.~Piccinini, A.~D.~Polosa, and V.~Riquer,
Phys. Rev. D \textbf{71}, 014028 (2005).

\bibitem{Bracco:2005kt}
M.~E.~Bracco, A.~Lozea, R.~D.~Matheus, F.~S.~Navarra, and M.~Nielsen,
Phys. Lett. B \textbf{624}, 217 (2005).

\bibitem{Browder:2003fk}
T.~E.~Browder, S.~Pakvasa, and A.~A.~Petrov,
Phys. Lett. B \textbf{578}, 365 (2004).

\bibitem{Lu:2006ry}
J.~Lu, X.~L.~Chen, W.~Z.~Deng, and S.~L.~Zhu,
Phys. Rev. D \textbf{73}, 054012 (2006).

\bibitem{Bicudo:2005de}
P.~Bicudo,
Phys. Rev. D \textbf{74}, 036008 (2006).

\bibitem{MartinezTorres:2011pr}
A.~Martinez Torres, L.~R.~Dai, C.~Koren, D.~Jido, and E.~Oset,
Phys. Rev. D \textbf{85}, 014027 (2012).

\bibitem{Mohler:2013rwa}
D.~Mohler, C.~B.~Lang, L.~Leskovec, S.~Prelovsek, and R.~M.~Woloshyn,
Phys. Rev. Lett. \textbf{111}, 222001 (2013).

\bibitem{Tang:2023yls}
M.~N.~Tang, Y.~H.~Lin, F.~K.~Guo, C.~Hanhart, and U.~G.~Mei\ss{}ner,
Commun. Theor. Phys. \textbf{75}, 055203 (2023).

\bibitem{BaBar:2003oey}
B.~Aubert \textit{et al.} (BaBar Collaboration),
Phys. Rev. Lett. \textbf{90}, 242001 (2003).

\bibitem{CLEO:2003ggt}
D.~Besson \textit{et al.} (CLEO Collaboration),
Phys. Rev. D \textbf{68}, 032002 (2003)
[erratum: Phys. Rev. D \textbf{75}, 119908 (2007)].

\bibitem{Belle:2003kup}
Y.~Mikami \textit{et al.} (Belle Collaboration),
Phys. Rev. Lett. \textbf{92}, 012002 (2004).

\bibitem{BESIII:2017vdm}
M.~Ablikim \textit{et al.} (BESIII Collaboration),
Phys. Rev. D \textbf{97}, 051103 (2018).

\bibitem{Chen:2022asf}
H.~X.~Chen, W.~Chen, X.~Liu, Y.~R.~Liu and S.~L.~Zhu,
Rept. Prog. Phys. \textbf{86}, 026201 (2023).

\bibitem{BaBar:2006eep}
B.~Aubert \textit{et al.} (BaBar Collaboration),
Phys. Rev. D \textbf{74}, 032007 (2006).

\bibitem{Faessler:2007gv}
A.~Faessler, T.~Gutsche, V.~E.~Lyubovitskij, and Y.~L.~Ma,
Phys. Rev. D \textbf{76}, 014005 (2007).

\bibitem{Fu:2021wde}
H.~L.~Fu, H.~W.~Grie\ss{}hammer, F.~K.~Guo, C.~Hanhart, and U.~G.~Mei\ss{}ner,
Eur. Phys. J. A \textbf{58}, 70 (2022).

\bibitem{Lutz:2007sk}
M.~F.~M.~Lutz and M.~Soyeur,
Nucl. Phys. A \textbf{813}, 14 (2008).

\bibitem{Godfrey:2003kg}
S.~Godfrey,
Phys. Lett. B \textbf{568}, 254 (2003).


\bibitem{Belle:2000cnh} A.~Abashian \textit{et al.} (Belle Collaboration), Nucl. Instrum. Meth. A \textbf{479}, 117 (2002).
\bibitem{PTEP_belle} J.~Brodzicka {\it et al.}, PTEP \textbf{2012}, 04D001 (2012).
\bibitem{Kurokawa:2001nw} S.~Kurokawa and E.~Kikutani, Nucl. Instrum. Meth. A \textbf{499}, 1 (2003). 
\bibitem{PTEP_kekb} T.~Abe {\it et al.}, PTEP \textbf{2013}, 03A001 (2013).
\bibitem{Belle-II:2010dht} T.~Abe \textit{et al.} (Belle~II Collaboration), arXiv:1011.0352.
\bibitem{Akai:2018mbz} K.~Akai \textit{et al.},
Nucl. Instrum. Meth. A \textbf{907}, 188 (2018).

\bibitem{basf2} T.~Kuhr, C.~Pulvermacher, M Ritter, T.~Hauth, and N.~Braun (Belle II Software Framework Group), Comput. Softw. Big Sci. \textbf{3}, 1 (2019).
\bibitem{basf2_repo} Belle II collaboration, Belle II Analysis Software Framework (basf2), \href{https://doi.org/10.5281/zenodo.5574115}{https://doi.org/10.5281/zenodo.5574115}.
\bibitem{Gelb:2018agf} M.~Gelb \textit{et al.}, Comput. Softw. Big Sci. \textbf{2}, 9 (2018).

\bibitem{Jadach:1999vf}
S.~Jadach, B.~F.~L.~Ward and Z.~Was,
Comput. Phys. Commun. \textbf{130}, 260 (2000).

\bibitem{pythia1} T.~Sj{\"{o}}strand {\it et al.}, Comput. Phys. Commun. {\bf 135}, 238 (2001).
\bibitem{pythia2} T.~Sj{\"{o}}strand {\it et al.}, Comput. Phys. Commun. {\bf 191}, 159 (2015).

\bibitem{PDG}
S.~Navas \textit{et al.} (Particle Data Group),
Phys. Rev. D \textbf{110}, 030001 (2024).
\bibitem{BESIII:2020ctr} M.~Ablikim \textit{et al.} (BESIII Collaboration),
Phys. Rev. D \textbf{104}, 012016 (2021).

\bibitem{geant3} R.~Brun {\it et al.}, GEANT 3: user's guide Geant 3.10, Geant 3.11, CERN Report No. DD/EE/84-1, 1984.
\bibitem{geant4} S.~Agostinelli \textit{et al.} (GEANT4 Collaboration), Nucl. Instrum. Meth. A \textbf{506}, 250 (2003).

\bibitem{pid} E.~Nakano, Nucl. Instrum. Meth. A {\bf 494}, 402 (2002).
\bibitem{Belle-II:2025tpe}
I.~Adachi \textit{et al.} (Belle II Collaboration),
Eur. Phys. J. C \textbf{85}, 1237 (2025).

\bibitem{Punzi:2003bu}
G.~Punzi,
eConf \textbf{C030908}, MODT002 (2003).

\bibitem{zhouxy_topo} X.~Zhou, S.~Du, G.~Li and C.~Shen, Comput. Phys. Commun. \textbf{258}, 107540 (2021).


\bibitem{CB_fun} 
M.~Oreglia,
SLAC report SLAC-0236 (1980).

\bibitem{significance} S. S. Wilks, Ann. Math. Stat. {\bf9}, 60 (1938).

\bibitem{Ke:2013zs}
H.~W.~Ke, X.~Q.~Li, and Y.~L.~Shi,
Phys. Rev. D \textbf{87}, 054022 (2013).
\end{thebibliography}
